\documentclass[12pt]{article}
\usepackage{amssymb}
\begin{document}
\setlength{\textheight}{9.4in}
\setlength{\topmargin}{-0.6in}
\setlength{\oddsidemargin}{0.15in}
\setlength{\evensidemargin}{0.4in}
\renewcommand{\thefootnote}{\fnsymbol{footnote}}
\newcommand{\e}{{\bf E}}
\newcommand{\m}{{\bf B}}
\newcommand{\h}{{\bf H}}
\renewcommand{\d}{{\bf D}}
\rightline{}

\vspace{1cm}

\begin{center}
{\Large {\bf Tubular D3-branes and their Dualities}}
\vspace{1truecm}

\large 
{S. Tamaryan,${}^{a,c}$\footnote[1]
{E-mail:sayat@moon.yerphi.am}
D. K. Park${}^{b}$\footnote[2]
{E-mail:dkpark@hep.kyungnam.ac.kr}
and H.J.W. M\"uller-Kirsten${}^c$\footnote[3]
{E-mail:mueller1@physik.uni-kl.de}}

\vspace{0.8cm}

\normalsize
${}^a${\it  Theory Department, Yerevan Physics Institute,\\ 
Yerevan-36, 375036, Armenia}\\
\vspace{0.4cm}
${}^b${\it Department of Physics, Kyungnam University,\\ 
Masan, 631-701, Korea}\\
\vspace{0.4cm}
${}^c${\it Department of  Physics, University of Kaiserslautern,\\ 
67653 Kaiserslautern, Germany}
\end{center}
\vspace{0.6cm}

{\centerline {\bf Abstract}}

\vspace{0.4cm}

A tubular D3-brane with electromagnetic flux is considered. It is verified 
that the quantized electric and magnetic charges are the F-string and D-string 
charges of the brane, respectively. The D3-brane with parallel electric 
and magnetic fields collapses to the bound state of strings. A D3-brane with 
nonzero Poynting vector can be viewed as expanded FD strings. A 
fundamental string is viewed as a collapsed D3-brane and is mapped to the 
$(n,m)$ strings via electric-magnetic duality rotations. This mapping is 
the same as a weak-strong duality transformation and therefore reveals the 
equivalence of the electric-magnetic and weak-strong dualities.

\newpage
\section{Introduction}
 
Recently supertubes,  as special tubular D2-branes, 
attracted considerable attention owing to
a previously unknown mechanism of stabilisation by
angular momentum as a result
of crossed constant electric and magnetic fields \cite{1}.
In particular it was shown in ref. \cite{1}
that the finite radius of the
tube results
from the product of F-string and D0-brane charges,
and the BPS energy is reached by a cancellation
of oppositely directed Poynting and centrifugal momenta,
thus requiring a nonzero angular momentum along the axis of the
supertube.
It was also shown that another supersymmetry
preserving configuration with the
same energy exists in the limit of vanishing
D2-brane radius with the D0-branes
aligning like beads along a IIA superstring,
with  vanishing angular momentum.
In these two cases the directions of the
$E$  and $B$  components 
of the electromagnetic field are reversed.
It is evident that such a seemingly simple model
immediately suggests its investigation
with respect to electric-magnetic \cite{2,3,4} 
 and weak-strong or S dualities \cite{5,6}.
Several subsequent investigations \cite{7,8,9} did not venture
in this direction, and this is therefore the main  objective  here.
Earlier  investigations \cite{10,11,12} 
studied the
stability  of branes in relation 
to their possible  collapse  to lower dimensional ones 
\cite{10,11,12} 
or the opposite, the expansion of branes into
higher dimensional ones \cite{9,13,14}
with further investigations in refs.\cite{14,15}.
  In view of
apparent physical shortcomings of the case of D2
 (e.g. the Poynting momentum
is a scalar),  it is natural to prefer the case
of D3 for the investigation since here
the connection between electric-magnetic and 
weak-strong dualities is exposed in the fullest physical
context, and at least weak-strong duality of this case
has been considered from time to time \cite{16}.  In fact we shall
show that weak-strong  duality may be understood
as the electric-magnetic duality of nonlinear 
electrodynamics.

Thus, in order to realise this idea, we proceed
in a number of definite steps. We first
demonstrate how quantised
electric and magnetic charges can be
connected with the numbers of
F-strings and D-strings dissolved in the
D3-brane.  We then show that the F-string is a D3-brane with
minimal electric charge.  In the third step
this D3-brane with minimal electric charge is mapped to the
D3-brane with arbitrary  electric and magnetic charges using
the electric-magnetic duality rotation.
Then, following this, we show that the D3-brane with arbitrary charges
is a composite string in the sense of Witten\cite{17} which can
also be described as an FD string\cite{18}.
Finally the
F-string
is mapped to this composite  string.

\section{The Action}

\noindent  

The D3-brane action is
\begin{equation}
I=-T_3\int d^4\xi\sqrt{-det(g_{\mu\nu}+2\pi F_{\mu\nu})},\qquad 
T_3=\frac{1}{(2\pi)^3g_s}, 
\label{bi1}
\end{equation}
where $g_{\mu\nu}$ is the induced metric tensor and 
$F_{\mu\nu}$ is the electromagnetic field strength tensor. The dilaton field 
is taken to be constant and target spacetime to be flat. We consider a 
D3-brane of cylindrical shape $\mathbb{R}^1\otimes \mathbb{S}^2$  which does 
not allow a variation of the  radius $r$ in time and along the axis. This 
choice is conditioned by our intention to investigate the brane's dualities.
The variable radius violates the electric-magnetic duality
invariance and therefore we keep it constant. 
The brane's worldvolume is parametrized in terms of the 
variables $(t,z, \theta , \varphi )$ and the cylindrical hypersurface is 
given by
\begin{eqnarray}
X^0=t,\quad X^1=z,\quad X^2=r\cos\theta,\qquad\qquad\qquad\nonumber\\
\quad X^3=r\sin\theta\cos\varphi,\quad X^4=r\sin\theta\sin\varphi,\quad 
others=constant.
\label{bi2}
\end{eqnarray}
The induced metric is
\begin{equation}
ds^2=-dt^2+dz^2+r^2d{\theta}^2+r^2{\sin}^2\theta\,d{\varphi}^2.
\label{bi3}
\end{equation}
Hereafter we use bold face letters for the 3-dimensional vectors and symbols 
$i,j,k$ for the space indices  $(z,\theta ,\varphi )$. The electric 
intensity $\e$ and magnetic induction $\m$ are defined by
\begin{equation}
{\bf E}_i=2\pi F_{0i},\qquad {\bf B}_i=2\pi\frac{{\epsilon}_{ijk}}{2}F^{jk}. 
\label{bi4}
\end{equation}
The Levi-Civita form ${\epsilon}_{ijk}$ in curved space is defined by
${\epsilon}_{ijk}=\sqrt{-g}{\varepsilon}_{ijk}$, where 
$-g\equiv det(-g_{\mu\nu})=r^4\sin^2\theta$ and ${\varepsilon}_{ijk}$ is 
the Levi-Civita form in flat space.

\noindent
The action reduces to
\begin{equation}
I=T_3\int {\cal L}\sqrt{-g}\,dtdzd\theta d\varphi,
\label{bi5}
\end{equation}
where the Lagrangian ${\cal L}$ is
\begin{equation}
{\cal L}=-\sqrt{1+{\m}^2-{\e}^2-(\e \cdot \m)^2},
\label{bi6}
\end{equation}
the brane tension is a normalization factor and the rest is the worldvolume 
invariant measure. We started from the D3-brane
action and obtained the Non-Linear Electrodynamic(NLE) action. The reason is 
that we suppressed all other degrees of freedom and kept only the gauge field.

The aspects of the NLE action we are interested in are most conveniently 
analyzed by introducing the electric induction $\d$,
\begin{equation}
\d=\frac{\partial {\cal L}}{\partial \e}=\frac{\e+\m (\e\cdot\m)}{\sqrt{1+
{\m}^2-{\e}^2-(\e\cdot \m)^2}}.
\label{bi7}
\end{equation}
Performing a Legendre transform we construct the Hamiltonian ${\cal H}$ 
as a function of $\m$ and $\d$,
\begin{equation}
{\cal H}=\sqrt{1+{\m}^2+{\d}^2+|\m\times\d|^2}.
\label{bi8}
\end{equation}
It has $SO(2)$ electric-magnetic duality rotation invariance
\begin{eqnarray}
\d & \rightarrow & \d\cos\alpha -\m\sin\alpha,\nonumber\\
\m & \rightarrow & \d\sin\alpha\, +\m\cos\alpha.
\label{bi9}
\end{eqnarray}
There is Legendre discrete duality at $\alpha =\pi /2$ which interchanges 
the electric and magnetic inductions.
The $SO(2)$ invariance of the Hamiltonian can be extended to 
$SL(2,\mathbb{R})$ invariance by including  a scalar dilaton field $\phi$ and 
a pseudo-scalar axion field $\chi$. For the time being we omit these fields 
for simplicity and return to this later. To complete this section we 
introduce the magnetic intensity $\h$ 
\begin{equation}
\h=-\frac{\partial {\cal L}}{\partial \m}=\frac{\m -\e (\e\cdot\m )}
{\sqrt{1+{\m}^2-{\e}^2-(\e\cdot\m)^2}}
\label{bi10}
\end{equation}
and write out the equations of motion in the static case:
\begin{equation}
{\varepsilon}^{ijk}\frac{\partial {\e}_i}{\partial x^j}=0,\qquad 
{\varepsilon}^{ijk}\frac{\partial {\h}_i}{\partial x^j}=0,
\label{bi11}
\end{equation}
\begin{equation}
\frac{\partial }{\partial x^k}\sqrt{-g}\,{\d}^k=0,\qquad 
\frac{\partial }{\partial x^k}\sqrt{-g}\,{\m}^k=0.
\label{bi12}
\end{equation}
The transformations (\ref{bi9}) can be equally described in terms of the 
intensities $\e$ and $\h$. They take the solution of the NLE action to
another solution. Note that the duality transformations (\ref{bi9}) are 
invariances of the field equations (\ref{bi11},\ref{bi12}) and Hamiltonian 
(\ref{bi8}) but not of the action (\ref{bi5}).

\section{Brane Charges}

\noindent

In this section we consider two static solutions of the NLE action 
(\ref{bi5}). In the first case electric and magnetic fields are parallel, 
the brane carries no angular momentum generated by crossed electric and 
magnetic fields and collapses to the bound state of the strings. In the second 
case electric and magnetic fields are perpendicular and the nonzero Poynting 
vector prevents the collapse of the brane. The flux in the D3-brane acts as 
a source for the strings. It comes 
from an integral over the whole world-volume and it is impossible to localise
the strings at a particular place in the brane's worldvolume. But they can 
appear as remnants when the cylindrical brane is squeezed to a line.

\noindent

Using Dirac charge quantization we express the 
energy of the solutions as a function of corresponding quantized electric and 
magnetic charges $n$ and $m$. In the limit of the vanishing radius $r$ 
of the brane, $r\rightarrow 0$, the term proportional to the area of the 
surface of the cylinder $(\sim r^2)$ in the expression of the energy  
can be neglected in leading order. Then the rest shows the strings smeared in 
the brane. In this way we establish that $n$ and $m$ are F-string and 
D-string charges of the brane, respectively .

\paragraph{Bound States of Strings}

\noindent

First we consider parallel uniform electric and uniform magnetic fields 
aligned with the $z$ axis, the 2-form field strength being given by
\begin{equation}
2\pi F=E\,dt\land dz+Br^2\sin\theta\,d\theta\land d\varphi.
\label{c1}
\end{equation}

The Dirac electric charge quantization condition is
\begin{equation}
n=\int\limits_{S^2}\frac{\delta I}{\delta F_{0i}}ds_i=
2\pi\int{\epsilon}_{ijk}\frac{\delta I}{\delta {\e}_i}dx^jdx^k,
\label{c2}
\end{equation}
where $n$ is an integer. The surface perpendicular to the electric field is 
the basesurface of the cylinder and is parametrized in terms of the variables 
$(\theta,\varphi)$. Then
\begin{equation}
T_3\int D\sqrt{-g}\,d\theta d\varphi=\frac{n}{2\pi};\qquad 
D={\d}^z.
\label{c4}
\end{equation}
The measure of the basesurface area $S_2$ is
\begin{equation}
dS_2=r^2\sin\theta\,d\theta d\varphi=\sqrt{-g}\,d\theta d\varphi,
\label{c5}
\end{equation}
so that eq.(\ref{c4}) takes the form
\begin{equation}
T_3\int DdS_2=\frac{n}{2\pi}. 
\label{c6}
\end{equation}
Eq.(\ref{c6}) implies that the electric induction flux through the basesurface is quantized, and as $D$ is constant it yields
\begin{equation}
D=\frac{\pi n g_s}{r^2}.
\label{c7}
\end{equation}
This result can be also derived from M-theory \cite{10,19}. 

\noindent

The magnetic flux quantization condition is
\begin{equation}
\int F_{\theta \varphi}\,d\theta d\varphi =2\pi m
\label{c8}
\end{equation}
where $m$ is an integer. It yields
\begin{equation}
B=\frac{\pi m}{r^2}.
\label{c9}
\end{equation}
The substitution of (\ref{c7}) and (\ref{c9}) into Hamiltonian (\ref{bi8}) 
and the integration over $\theta ,\varphi$ gives for the energy ${\cal E}$ 
\begin{equation}
{\cal E}=\sqrt{(T_3S_2)^2+(nT_s)^2+(mT_1)^2}\int dz.
\label{c10}
\end{equation}
where $T_s=1/2\pi$ and $T_1=1/2\pi g_s$ are the tensions of the F-string 
and D-string, respectively.
To uncover the dissolved strings we consider the energy ${\cal E}$ in the 
limit $S_2\rightarrow 0$. In pure electric case it can be presented as the 
mass of $n$ fundamental strings
\begin{equation}
{\cal E}(m=0)_{r\rightarrow 0}= nT_s\int dz.
\label{c11}
\end{equation}
From this asymptotoic relation we conclude that there are $n$ fundamental 
strings dissolved in the worldvolume of the D3-brane, and $n$ is 
the F-string charge of the D3-brane. In the pure magnetic case the energy 
becomes the mass of $m$ D-strings
\begin{equation}
{\cal E}(n=0)_{r\rightarrow 0}= mT_1\int dz,
\label{c12}
\end{equation}
with the obvious interpretation that there are $m$ D-strings dissolved in 
the D3-brane, and $m$ is the D-string charge of the D3-brane. 

\noindent

The energy (\ref{c10}) is minimal when $r=0$, as it should be. The brane 
shrinks to a line with the energy
\begin{equation}
{\cal E}=\sqrt{(nT_s)^2+(mT_1)^2}\int dz.
\label{c16}
\end{equation}
This is the bound state of $(n,m)$ strings. The D3-brane with charges $(n,m)$ 
and with no net D3-brane charge collapses to the $(n,m)$ strings and 
therefore the latter can be considered as a collapsed D3-brane. The mass 
formula (\ref{c16}) for the $SL(2,\mathbb{Z})$ multiplet was derived in 
\cite{17,20}. 
The basic condition was the fact that the string solutions are BPS-saturating 
states and preserve half of supersymmetry. Here the same relation is derived 
from the viewpoint of the supersymmetric D3-brane. The main point is that the 
string bound state is considered as a collapsed superbrane. This 
method allows to apply Legendre discrete duality to the $(n, m)$ strings. 
The latter interchanges the S-dual partners, i.e. F-strings and D-strings, 
resulting in $(m,n)$ strings.
\paragraph{Expanded FD strings}

\noindent

Eq.(\ref{bi11},\ref{bi12}) have nontrivial solutions with the perpendicular 
electric and magnetic fields. Suppose we have the electric field along the $z$ 
axis and the magnetic field along meridians, so only 
${\e}_z,{\d}_z,{\m}_{\theta},{\h}_{\theta}$ components are nonzero.  
The eq.(\ref{bi12}) gives a simple solution for ${\m}_{\theta}$
\begin{equation} 
{\m}_{\theta}=\frac{B}{\sin\theta},
\label{p1}
\end{equation}
where $B$ is an integration constant. The ${\e}_z$ components can be 
expressed via ${\d}_z$ and ${\m}_{\theta}$
\begin{equation} 
\e^2_z=\bigg(1+\frac{B^2}{r^2\sin^2\theta}\bigg)
\frac{\d^2_z}{1+\d^2_z}.
\label{p2}
\end{equation}
From the eq.(\ref{bi11}) it follows that ${\e}_z$ can not depend on the
variable $\theta$. The r.h.s of the eq.(\ref{p2}) satisfies this 
restriction only when
\begin{equation} 
{\d}_z=D\sin\theta,
\label{p3}
\end{equation}
where the constant $D$ is defined by
\begin{equation} 
D^2B^2=r^2.
\label{p4}
\end{equation}
Then eq.(\ref{p2}) becomes $\e_z^2=1$ and we rewrite it in the invariant 
form
\begin{equation}
{\e}^2=1.
\label{p5}
\end{equation}
This is the critical value of the electric intensity which forces the critical 
value of the magnetic intensity ${\h}^2=1$. Now we can make $SO(2)$ duality 
rotations and obtain more general solutions. As $\e$ and $\h$ are unit mutually perpendicular vectors, they will stay such after rotations (\ref{bi9}).
Consequently the condition (\ref{p5}) is right for any rotated solution and 
defines the true minimum of the NLE action. It has another interesting 
interpretation. The question whether the tubular D2-brane with the nonzero 
Poynting vector is supersymmetric has been clarified \cite{1}. 
The condition for the preservation of 1/4 of the supersymmetry was reduced to
the eq.(\ref{p5}). We come to the conclusion that the classical minimum 
of the NLE action is a BPS-saturating state.

\noindent 

The charge quantization requires the compact length of the variable $z$. 
Therefore we assume that the $z$ direction is compactified by a circle with 
the circumference $l$. We denote by $R$ the compactification radius, then 
$l=2\pi R$. 
The quantization repeats the procedure of the previous case 
resulting in
\begin{equation}
D=\frac{4ng_s}{r^2},\qquad B=\frac{m}{R}.
\label{p6}
\end{equation}
The energy expression takes the form
\begin{equation}
{\cal E}=\int \sqrt{(rT_3S_2^{\prime})^2+(4nT_sR)^2\sin^2\theta+
\frac{(mT_1r)^2}{\sin^2\theta}+\frac{4n^2m^2}{\pi^2r^2}}\,\sin\theta\,d\theta.
\label{p7}
\end{equation}
where $S_2^{\prime}=2\pi rl$ is the azimuthal lateral area of the brane.
The first term under the square root is the energy of the brane with no 
electromagnetic flux. The second term presents $n$ dissolved 
F-strings which is easy to see by setting $m=0$ and taking the limit 
$r\rightarrow 0$. 
\begin{equation}
{\cal E}(m=0)_{r\rightarrow 0}=nT_sl.
\label{p8}
\end{equation}
In the pure magnetic case the energy is
\begin{equation} 
{\cal E}(n=0)=mT_1\int \sqrt{\bigg(\frac{Rr^2}{m}\bigg)^2\sin^2\theta +r^2}
\,d\theta
\label{p9}
\end{equation}
In the limit $r\rightarrow 0$ this becomes
\begin{equation}
{\cal E}(n=0)_{r\rightarrow 0}=mT_1(\pi r).
\label{p10}
\end{equation}
This is a bunch of m D-strings with the length of a semicircle. Therefore $m$ 
is the brane's D-string charge. The integrand in (\ref{p9}) represents the 
length measure of an elliptic arc along which the energy density is 
distributed and the ellipse is mapped into the azimuthal lateral face of 
$\mathbb{R}^1\otimes \mathbb{S}^2$ becoming 
an elliptic helix with a pitch $r$.

The last term in (\ref{p7}) is generated by the 
Poynting vector and owing to it the energy grows infinitely 
when $r\rightarrow 0$. The nonzero Poynting vector $P(P\sim nm)$ 
stabilizes the brane at nonzero radius $r_0$. The substitution of 
(\ref{p6}) into  (\ref{p4}) determines the value of the radius $r_0$  
\begin{equation}
{r_0}^3=\frac{4\pi g_s}{R}\,|nm|.
\label{p11}
\end{equation}
The brane tension forces the collapse of the brane while the momentum 
works oppositely. The same value $r_0$ can be determined from the balance 
condition of these two acts. Indeed, equating the first and last 
terms under the square root (\ref{p7}) 
\begin{equation}
rT_3S_2^{\prime}=\frac{2|nm|}{\pi r}
\label{p12}
\end{equation}
yields the eq.(\ref{p11}).
Therefore the D3-brane with the given charges
can be regarded as an expanded FD string which is 
self-supported from collapse by the momentum $\d\times\m$.

\noindent

Legendre discrete transformation maps the above D3-brane to the its 
electromagnetic dual D3-brane
\begin{equation}
\d_{\theta}=\frac{D}{\sin\theta},\quad \m_z=B\sin\theta ;\qquad |BD|=r.
\label{p13}
\end{equation}
The quantization of the electromagnetic charges gives rise to the S-dual 
picture, i.e. D-strings are aligned along the cylinder axis and F-strings 
are wound around them helically.
One should note that the precise form of the helix does not have 
any meaning because the worldvolume reparametrization invariance allows to
transmute it to another form. What is definite, is that F and D strings are
wound around each other and form a 3-dimensional braid. Necessarily they 
carry nonzero angular momentum, otherwise the system collapses to the bound 
state of the strings.

\section{Dualities}

\noindent

The weak-strong duality may be realized as a NLE electric-magnetic duality 
on the D3-brane worldvolume theory. In this section we envisage the 
possibility that this conception is presented in the precise form. In the 
case of constant flux in the {\it flat direction} of the tubular D3-brane 
we explicitly demonstrate the equivalence of these dualities.

\noindent

Equations of motion of the NLE action (\ref{bi5}) and Hamiltonian (\ref{bi8}) 
are invariant under $SO(2)$ electric-magnetic duality rotations. It has been 
shown that this invariance becomes $SL(2,\mathbb{R})$ symmetry by including 
dilaton and axion fields \cite{2,3,4}. The Lagrangian takes the form 
\cite{3,5,6} 
\begin{equation}
{\cal L}^{\prime}=e^{-\phi}{\cal L}+\chi(\e\cdot\m).
\label{d1}
\end{equation}
It defines the electric induction ${\d}^{\prime}$
\begin{equation}
{\d}^{\prime}=e^{-\phi}\d+\chi\m
\label{d2}
\end{equation}
which shows that the axion charge of the string is provided by electric flux
in the worldvolume of the brane. To restore the symmetric dependence of the 
Hamiltonian on the inductions ${\d}^{\prime}$ and $\h$ one has also to 
rescale the metric tensor.

\noindent

The duality transformations are conveniently described in terms of the 
complex field $\lambda=\chi+ie^{-\phi}$. They mix electric and magnetic fields 
\cite{3}
\begin{eqnarray}
\left(\begin{array}{c}
{\d}^{\prime}\\
\m
\end{array}\right)
\rightarrow S
\left(\begin{array}{c}
{\d}^{\prime}\\
\m
\end{array}\right)
\label{d3}
\end{eqnarray}
and induce a M\"obius transformation for the $\lambda$ field,
\begin{equation}
\lambda\rightarrow\frac{a\lambda +b}{c\lambda+d},
\label{d4}
\end{equation}
where $S\in SL(2,\mathbb{R})$ is
\begin{eqnarray}
S=
\left(\begin{array}{cc}
a & b\\
c & d
\end{array}\right)
\label{d5}
\end{eqnarray}
with $ad-bc=1$.

\noindent

In the static case ${\d}^{\prime}$ satisfies the same Gauss law and 
quantization condition (\ref{c2}). Its normalization change is caused by
the replacement $1/g_s\rightarrow exp(-\phi)$. For instance, 
instead of eq.(\ref{c7}) one gets
\begin{equation}
D^{\prime}=\frac{\pi n}{r^2}.
\label{d6}
\end{equation}
Owing to this, $SL(2,\mathbb{Z})$ transformations can be formulated in 
terms of $\d$ and $\m$ introduced in the previous section
\begin{eqnarray}
\left(\begin{array}{c}
\frac{1}{g_s}\d\\
\m
\end{array}\right)
\rightarrow S
\left(\begin{array}{c}
\frac{1}{g_s}\d\\
\m
\end{array}\right)
\label{d7}
\end{eqnarray}
 for the quantized charges.  
We now consider a D3-brane with minimal electric charge, $n=1$. This has to 
shrink to the F-string and we are actually considering a fundamental string.
We will map it to the $(n,m)$ strings for any relatively prime pair
$n$ and $m$. Indeed, it is well known that the Diophante equation
\begin{equation}
nx-my=1
\label{d8}
\end{equation}
always has a solution for the such a pair of integers. In fact, it has  
countably many solutions, because if $(x=i,y=j)$ is a solution then for any 
integer $k$ $(x=km+i,y=kn+j)$ is also a solution. Then for the given pair  
$(n,m)$ we have a matrix $S\in SL(2,\mathbb{Z})$
\begin{eqnarray}
S=
\left(\begin{array}{cc}
n & j\\
m & i
\end{array}\right).
\label{d9}
\end{eqnarray}
We perform an electric-magnetic duality rotation (\ref{d7}) via $S$, i.e.
\begin{eqnarray}
\left(\begin{array}{cc}
n & j\\
m & i
\end{array}\right)
\frac{\pi}{r^2}
\left(\begin{array}{c}
1\\
0
\end{array}\right)
=\frac{\pi }{r^2}
\left(\begin{array}{c}
n\\
m
\end{array}\right).
\label{d10}
\end{eqnarray}
It produced parallel electric and magnetic fields with corresponding charges 
$n$ and $m$. We already ascertained that the D3-brane with these charges 
collapses to the bound state of strings. Thus electric-magnetic duality 
mapped the fundamental string to the $(n,m)$ strings which is a 
weak-strong duality transformation \cite{17}.

\noindent

We note that eq.(\ref{d8}) has no solutions when $n$ and $m$ are not 
relatively prime. Consequently we can not map the fundamental string to the 
FD strings in this case, they are not in the same $SL(2,\mathbb{Z})$ 
multiplet. This means that the bound state of strings becomes unstable when 
$n$ and $m$ have a common multiplier.   

\section{Conclusions}

\noindent

The NLE static equations (\ref{bi11},\ref{bi12}) admit a wider class of 
solutions 
with nonzero Poynting vector, but we did not consider these. Our justification 
is that these solutions do not allow the application of $SL(2,\mathbb{Z})$ 
tranformations. In curved subspaces SO(2) duality is maintained but other 
dualities, 
in general, are broken \cite{4}. One can see this from the expression 
(\ref{p6}), the charges are quantized in different units and they can not
be mapped to each other. There are two possibilities to get D3-brane charges 
quantized in the same units, that is to consider branes of the shape 
$\mathbb{R}^2\otimes\mathbb{S}^1$ or
$\mathbb{R}^1\otimes\mathbb{S}^1\otimes\mathbb{S}^1$. In both cases there are 
two orthogonal equivalent directions along which the electric and magnetic 
fields can be directed. It will restore $SL(2,\mathbb{R})$ electric-magnetic 
duality and describe an $SL(2,\mathbb{Z})$ multiplet of D3-branes which can be 
mapped to each other.

\smallskip

{\bf Acknowledgments:}
\newline
S.T. acknowledges support by the Deutsche Forschungsgemeinschaft(DFG).
D.K.P. was supported by grant R05-2001-000-00106-0 from the Basic Research 
Program of the Korea Science and Engineering Foundation.

\end{document}